\documentclass[prd, twocolumn, nofootinbib, floatfix]{revtex4-2} 

\usepackage{amsmath}
\usepackage{graphicx}
\usepackage{dcolumn}
\usepackage{bm}
\usepackage{epsfig}
\usepackage{amssymb,latexsym,mathrsfs,empheq}
\usepackage{graphicx}
\usepackage{color}
\usepackage{xcolor} 
\usepackage{mathtools} 
\usepackage{hyperref} 
\usepackage{cleveref}

\hypersetup{
    colorlinks=true,
    linkcolor=red,
    citecolor=blue,
} 

\newcommand{\be}{\begin{equation}}
\newcommand{\ee}{\end{equation}}
\newcommand{\bs}{\begin{split}} 
\newcommand{\bea}{\begin{eqnarray}}
\newcommand{\eea}{\end{eqnarray}}

\newcommand{\ode}{\Omega_{\rm de}}
\newcommand{\odet}{\Omega_{{\rm de},0}} 
\newcommand{\wde}{w_{\rm de}} 
\newcommand{\wdef}{w_{\rm de}^{\rm eff}}
\newcommand{\wcf}{w_m^{\rm eff}}
\newcommand{\rde}{\rho_{\rm de}} 
\newcommand{\rdef}{\rho_{\rm de}^{\rm eff}}
\newcommand{\rcf}{\rho_m^{\rm eff}}
\newcommand{\lcdm}{$\Lambda$CDM}

\newcommand{\mss}{M_\star^2} 
\newcommand{\mpl}{M_{\rm Pl}^2}
\newcommand{\alm}{\alpha_M}

\begin{document}

\title{Uplifting, Depressing, and Tilting Dark Energy} 

\author{Eric V.\ Linder} 
\affiliation{
Berkeley Center for Cosmological Physics \& Berkeley Lab, 
University of California, Berkeley, CA 94720, USA
} 

\begin{abstract} 
Current data in the form of baryon acoustic oscillation, 
supernova, and cosmic microwave background distances prefer 
a cosmology that accelerates more strongly than \lcdm\ at 
$z\approx0.5$--1.5, and more weakly at $z\lesssim0.5$. We 
examine dark energy physics that can accommodate this, showing 
that interactions (decays, coupling to matter, nonminimal 
coupling to gravity) fairly generically tend not to give a 
satisfactory solution 
(in terms of fitting both distances and growth) 
even if they enable the effective dark 
energy equation of state to cross $w=-1$. To fit the 
cosmological data it appears the dark energy by itself must cross $w=-1$, 
a highly unusual physical behavior.  
\end{abstract} 

\date{\today} 

\maketitle

\section{Introduction} 

After several decades in which a cosmological constant $\Lambda$ 
(along with baryonic matter, cold dark matter, and primordial 
radiation) provided a satisfactory fit to cosmological 
observations (if lacking a clear theoretical foundation), we 
are now at a time when cosmic distance probes prefer a dynamical 
dark energy \cite{dr2,dr2lya}, moreover one that gives stronger acceleration than 
$\Lambda$ at mid redshifts and weaker acceleration closer to the 
present. 

Any two probes of baryon acoustic oscillation (BAO) distances, 
supernova (SN) distances, or cosmic microwave background (CMB) 
observations (with leverage dominated by the distance to last 
scattering) indicate this at the $\sim2\sigma$ or $3\sigma$ level, while all 
three probes together do so at the $\sim3-4\sigma$ level. That 
does not attain the desirable $5\sigma$ criterion often used in 
physics, but the potential physics indicated if the data are 
taken at face value is worth exploring. 

Indeed, the behavior required is not what was previously 
expected on the basis on simple, apparently robust physics. 
The competition between cosmic expansion and dynamics in an 
effective field potential should guide the dark energy into 
either the thawing (moving away from $\Lambda$'s static 
equation of state $w=-1$) or freezing (moving toward $w=-1$) 
classes \cite{caldlin,paths}. Other regions of the dynamics 
phase space were ``zones of avoidance'', where some standard physical 
principle needed to be violated. 

In \cite{away} the zones of avoidance, and the principles 
violated, were discussed in detail. The conclusion was 
that to satisfy all the required properties, especially that 
the dynamics had to cross $w=-1$, the dark energy 
generally had to interact in some manner. Here we examine the 
possibilities, while keeping to general characteristics without 
assuming any specific model, 
and assess whether such dynamics can be 
achieved in a fairly straightforward manner. 

Section~\ref{sec:props} reviews the properties required by 
the data and how this connects with the zone of avoidance 
physical principles. The effects of interaction on the effective equations of state, cosmic acceleration, 
and growth of large scale structure, 
are outlined in Section~\ref{sec:intx}. In Section~\ref{sec:up} 
we consider phantom dark energy 
(e.g.\ emerging from a phase transition) that 
is uplifted through interactions to cross $w=-1$. 
Section~\ref{sec:down} examines depression of conventional 
thawing dark energy through interactions 
to cross $w=-1$. Having both effects in turn, tilting the dark energy to 
cross $w=-1$, is 
discussed in Section~\ref{sec:tilt}. 
We conclude in Section~\ref{sec:concl}.

\section{Data and Dynamics} \label{sec:props} 

The scalar field equation of motion (Klein-Gordon equation) 
for the (effective) dark energy governs the field dynamics 
through a competition between a Hubble friction term due 
to cosmic expansion and an effective potential/pressure 
steepness with respect to the field. (See for example 
Appendix~A of \cite{bellsaw} for the effective dark energy 
pressure gradient $P_\phi$ for the general Horndeski 
scalar-tensor theory.) At various cosmic epochs, one effect will 
dominate over the other, unless there is some fine tuning. 
This leads to the thawing and freezing regimes, and for the 
canonical Klein-Gordon equation these are 
restricted to well localized regions of the equation of state 
phase space \cite{caldlin,paths}. 

Yet the current data lies outside of either region.  
Ref.~\cite{away} identified the best fit of current data as 
lying in the zone of avoidance with greater time variation 
$w'$ than allowed by the thawing class. The physical principle 
behind the avoidance of this region of phase space is a simple 
one: the dark energy evolved in an old universe that had many 
efolds of matter (and radiation) domination, where the Hubble 
friction was high, so $w'$ {\it could not\/} thaw so rapidly. 
The solution proposed by \cite{away} was to keep matter domination 
-- since it is so essential to, and successful for, primordial 
nucleosynthesis, CMB, and growth of large scale structure -- 
but have dark energy ``appear'' only at redshift $z\approx1-2$, through 
a phase transition. Thus this gives a loophole in this zone 
of avoidance. 

If dark energy appears and grows in energy density, then 
its effective equation of state $w<-1$ (sometimes called the 
phantom regime). This will give 
acceleration stronger than from $\Lambda$. The data also indicates, 
however, that more recently the acceleration is weaker than 
from $\Lambda$, i.e.\ $w>-1$, so we need a mechanism by which 
dark energy can cross $w=-1$, sometimes called the phantom 
divide. Ref.~\cite{away} summarized the four properties of 
dark energy the data points to: 
\begin{itemize} 
\item Having $w<-1$ at $z\gtrsim0.5$, 
\item Superevolving faster than matter dominated Hubble friction 
should allow, 
\item Crossing $w=-1$, 
\item Having $w>-1$ at $z\lesssim0.5$. 
\end{itemize} 

The first two properties indeed can be characteristic of 
a phase transition, but the second two require some additional 
element. As mentioned in \cite{away}, a natural possibility 
is some interaction to drive the field across $w=-1$, rather 
going to an attractor such as de Sitter $w=-1$. Note that 
without interaction of some sort, even noncanonical kinetic 
terms cannot bring dark energy across $w=-1$ 
\cite{vikman,calddor,sen}. 
While one could 
do this with multiple scalar fields, this increases fine tuning 
issues -- requiring one field with $w<-1$, one with $w>-1$, 
and special timing and amplitude between the two. Here we 
focus on various classes of interaction involving a single 
scalar field. 
We emphasize that we do not assume specific models, 
but can still find quite general conclusions. For 
some recent examples of specific interaction models see 
\cite{2503.10806,2503.16415,2504.07679,2505.10410,2505.24732}.

\section{Interactions} \label{sec:intx} 

Interactions between dark energy and other components shift 
the effective equation of state of each component from its 
intrinsic, or ``bare'', value. Thus the crossing of $w=-1$ 
could arise from a phantom equation of state being somewhat 
uplifted, or a bare $w>-1$ equation of state being depressed.  
But there are aspects, and consequences, that must be taken 
into account. 

First, whatever the dark energy is coupled to also 
changes, which will impact probes beyond distances. There are 
certainly stringent limits on interactions with baryons and 
photons, less so on neutrinos (but the 
energy density in neutrinos is so small that it seems unlikely to 
cause a large effect). Dark matter is less constrained, 
but since it plays such a significant role in the growth of 
large scale structure (including the integrated Sachs-Wolfe 
effect), even modest shifts from being pressureless, 
i.e.\ $w_{\rm dm}=0$, tend to make such interactions 
unviable observationally.  

Second, specifying the effect on the energy 
density is not the whole picture, one should also specify 
the impact on momentum, e.g.\ when one component decays 
into or scatters off another. If one starts from an action 
then one can include the momentum and perturbation effects, 
otherwise they may be ad hoc or neglected. Here we deal with 
only distance probes in order to explore what types of 
interactions {\it might\/} work for cosmic expansion, and 
so can consider only the energy density, but 
if a model looks promising it should be redeveloped from 
an action and the consequences on other probes considered. 

We begin generally, by considering an arbitrary interaction 
$Q$ between dark energy and some component such as dark matter, 
\bea 
\dot\rde+3H(1+\wde)\rde&=&-Q\\ 
\dot\rho_m+3H\rho_m&=&Q \ . 
\eea 
Here, we take the bare (noninteracting) dark energy equation 
of state to be $\wde$ and the bare dark matter equation of 
state to be zero. We can easily rewrite these equations to 
define effective equations of state for each component in  
the presence of the interaction, with 
\bea 
\wdef&=&\wde+\frac{Q}{3H\rde}\\ 
\wcf&=&\frac{-Q}{3H\rho_m}\ . 
\eea 
Since near the present $\rde\approx\rho_m$, then however 
much we shift $\wde$ up (if $Q>0$) then we shift $w_m$ down. 
This follows directly from conservation of the total energy 
momentum. 
That is why considering consequences on growth, for example, 
is so crucial -- if we want to go from $\wde=-1.2$, say, 
to $\wdef=-0.7$, then we {\it must\/} have $\wcf\approx-0.5$. That is 
a huge shift that will dramatically impact large scale 
structure growth. Recall that we have not had to specify 
the form of $Q$ for this conclusion. 

An interesting, and in some sense simpler and more intuitive, way of viewing 
the impact of interactions on the cosmic expansion is to 
look directly at the deceleration parameter $q(z)$. Recall 
that 
\be 
q=-1+\frac{\dot H}{H^2} \ , 
\ee 
from which we see 
\bea 
q(z)&=&\frac{1}{2}\sum_i (1+3w_i)\Omega_i(z)\\ 
&=&\frac{1}{2}\left[1+3\wcf\Omega_m(z)+3\wdef\Omega_{\rm de}(z)\right] \label{eq:q2}\\ 
&=&\frac{1}{2}\left[1-\frac{Q}{H}\frac{8\pi G}{3H^2}+3\wde\Omega_{\rm de}+\frac{Q}{H}\frac{8\pi G}{3H^2}\right]\\ 
&=&\frac{1}{2}\left[1+3\wde\Omega_{\rm de}(z)\right]\ . \label{eq:q3} 
\eea 

This has two interesting implications. First, the interaction 
$Q$ does not explicitly appear! It is hidden inside the evolution 
of the effective dark energy density as a fraction of the critical density, 
$\Omega_{\rm de}(z)$, while otherwise only the {\it bare\/} dark 
energy equation of state $\wde$ enters. Second, if we consider the 
acceleration $q$ at a certain epoch, say today, then 
$q_0=(1/2)[1+3w_0\Omega_{{\rm de},0}]$. The data can fit for 
$\Omega_{{\rm de},0}$ and $q_0$, and we can determine what 
$w_0\equiv w_{{\rm de},0}$ 
(the bare quantity before interaction, not the ``measured'' 
$w_{{\rm de},0}^{\rm eff}$) must be 
to fit the data. 

In Figure~9 of \cite{dr2de} this is done through a Gaussian 
process applied to the data, without any function assumed for 
$\wde(z)$ or other such quantities. The result is $q_0\approx-0.3\pm0.1$. 
To get a theory estimate for $\Omega_{{\rm de},0}$, which recall is 
the effective not bare dark energy density, one would have to 
compute it for some specific interaction $Q(a)$, but it seems 
quite reasonable to assume $\Omega_{{\rm de},0}\approx0.7\pm0.1$ 
so that large scale structure growth is not too strongly 
affected. This would imply that the bare dark energy equation 
of state is $w_0\approx-0.76$. (Note that we expect $q_0$ and 
$\Omega_{{\rm de},0}$ to be anticorrelated, so the more 
negative values of $q_0$ correspond to the larger values of 
$\Omega_{{\rm de},0}$; $(q_0,\Omega_{{\rm de},0})=(-0.3,0.7)$, 
$(-0.4,0.8)$, $(-0.2,0.6)$ respectively gives $w_0=-0.76$, 
$-0.75$, $-0.78$.) 

However, Figure~9 of \cite{dr2de} also shows that the data 
requires cosmic acceleration to be stronger than in \lcdm\ at 
$z\approx0.5-1.5$, the sign of $w<-1$. Thus, we are squeezed 
between stronger acceleration at 
higher redshift and weaker acceleration at lower 
redshift. Dark energy 
interaction as a mechanism to cross $w=-1$ appears moot: the 
bare dark energy itself must already have the property of 
crossing $w=-1$ in order to accord with the data.

\section{Uplifting Phantoms} \label{sec:up} 

To go into a bit more detail, consider the case where the 
dark energy intrinsically is always phantom, $\wde(z)\le-1$. 
This can account for the stronger acceleration at $z\approx0.5-1$, 
but to achieve the weaker acceleration at $z\lesssim0.5$, e.g.\ 
$q_0=-0.3\pm0.1$, we need to uplift the dark energy to cross 
$w=-1$. Can this be done viably? 

We can rewrite the acceleration parameter measurement today 
in terms of the required effective dark energy density today, 
given that $w_0\lesssim-1$, 
\be 
\odet= \frac{2q_0-1}{3w_0}\le 0.53\pm0.07\ . \label{eq:odet} 
\ee 
This seems highly unlikely to be viable, given that it 
requires such a significant interaction to lower $\odet$ 
(and raise $\Omega_{m,0}$); this would impact growth of 
large scale structure significantly. 

In particular, the dark matter equation of state changes, 
being no longer effectively pressureless. Rewriting 
Eqs.~(\ref{eq:q2}) and (\ref{eq:q3}), 
\be 
\wcf=\frac{\Omega_{\rm de}}{\Omega_m}\,\left(\wde-\wdef\right)\ . 
\ee 
Using the limit on $\odet$ from above (assuming the most 
liberal limit $w_0=-1$), and taking $\wdef=-0.8\pm0.07$ from 
the Gaussian process reconstruction of Fig.~9 in \cite{dr2de}, 
yields $\wcf=-0.23$. Again taking the extreme values with 
their expected anticorrelation, $(\odet,\wdef)=(0.6,-0.87)$ 
and $(0.47,-0.73)$ respectively produces 
$\wcf=-0.20$ and $-0.24$. Observationally derived limits on 
$\wcf$ vary considerably depending on the time dependence 
of $\wcf$ -- ranging from the $10^{-3}$ level for the time 
independent case (mostly from the CMB including the ISW effect) to the $10^{-1}$ 
level for independent bins in redshift (mostly from growth of structure); 
for some classic and recent papers see e.g.\ 
\cite{0410621,1309.6971,1601.05097,2004.09572,2307.05155,2307.09522,2308.13617,2311.13795}. However, these generally 
underestimate the effect since they hold $\Omega_m\sim0.3$, 
not the $\Omega_m\sim0.5$ required above. A more thorough 
calculation would need to assume a specific form of 
interaction, which we have avoided here. 

The general lesson is that 
\be 
\wcf \approx \wde-\wdef \ , \label{eq:dwc} 
\ee 
so if we want to uplift phantom dark energy then we must 
make negative the dark matter equation of state by 
approximately the same amount. While we have taken the 
most benign limit above, $w_0=-1$, in fact many phantom 
dark energy cases have the intrinsic value $w_0\approx[-1.1-1.3]$ 
(since the universe today is not fully dark energy dominated and 
hence not at a de Sitter attractor), 
thus requiring $\wcf\approx [-0.3,-0.6]$. 
Uplifting dark energy seems quite difficult to make 
viable.

\section{Depressing Dark Energy} \label{sec:down} 

Taking dark energy with intrinsic $w>-1$, perhaps even a 
quintessence model, and pulling it down across $w=-1$ 
through some interaction is another possibility. However, 
this runs into many similar problems. 

To get the stronger acceleration at $z\approx 0.5-1.5$, 
we must have $\wdef<-1$ there. This initially sounds 
attractive, in that the thawing class of dark energy 
has $w=-1$ at high redshift, and so needs relatively less 
pull to bring it below $w=-1$. However, this does require 
the interaction to be significant at these high redshifts, 
exacerbating the issue of shifting the dark matter equation 
of state (now to values greater than zero) at higher redshift 
where there is a greater lever arm affecting growth (and 
the ISW effect). And of course if the interaction persists 
to even higher redshift (for example because $Q$ is 
proportional to $\rho_m$ as in dark matter decay models), 
we run into the highly sensitive CMB constraints. 
Finally, to achieve $\wdef\approx-0.8$ today, one requires 
the intrinsic $w_0$ to be even less negative, needing a 
very strong thawer (which tends to be fine tuned, i.e.\ 
the field must be nearly motionless for a long time so 
it stays near $w=-1$, 
then rapidly dive down an exceptionally steep potential). 

Let us carry out for depressed dark energy 
a similar analysis as for uplifted dark 
energy, this time using that the acceleration is stronger 
than in \lcdm\ at $z=1$. The Gaussian process reconstruction 
in Fig.~9 of \cite{dr2de} gives $q_1\equiv q(z=1)\approx 0.1\pm0.03$. 
Allowing $\ode(z=1)$ to deviate by up to 10\% from its \lcdm\ 
value (which would impart a significant change to growth of 
structure), we find $(q_1,\ode(z=1))=(0.13,0.20)$, $(0.1,0.23)$, 
$(0.07,0.25)$ leads to $w_1\equiv \wde(z=1)=-1.19$, $-1.16$, 
and $-1.13$ respectively. Thus again the data seems to 
require that the intrinsic dark energy behavior, before any 
interaction, must cross $w=-1$, therefore removing the main 
motivation for adding interactions. 

If we force $w_1\ge-1$, then this requires 
$\ode(z=1)\ge 0.27\pm0.02$, some 20\% higher than in the 
\lcdm\ case. This will have a concomitant impact on growth 
of structure. However the dark matter equation of state is 
not as severely affected as in the uplifted case. 
Equation~(\ref{eq:dwc}) still holds, so for $\wdef$ depressed 
by $\sim0.2$ below $\wde$ at $z=1$, we get 
$\wcf(z=1)\approx 0.06$. 

At $z=0$, where the acceleration is weaker than in \lcdm, 
the Gaussian process reconstruction gives $\wdef\approx-0.8$ 
so taking into account the depression the intrinsic  
$w_0\gtrsim-0.6$, say. Using Eq.~(\ref{eq:odet}) this 
would force $\odet\gtrsim0.89$, again causing significant 
impact on growth. If there were somehow no depression in 
the dark energy equation of state at $z=0$, then 
$\odet\gtrsim0.75$, taking into account the anticorrelation 
of $\odet$ with $q_0$. 

Thus, while depressing dark energy has some advantages 
over uplifting dark energy, it still poses challenges to 
being viable.

\section{Tilting Dark Energy} \label{sec:tilt} 

The common problem between uplifting dark energy and 
depressing dark energy is that the shifts to bring an  
intrinsic phantom model up enough to satisfy the low 
redshift weaker acceleration, or an intrinsically $w>-1$ 
model down enough to satisfy the high redshift stronger 
acceleration, are so large that the matter sector, and 
growth of structure, is severely affected. To avoid this 
the dark energy has to already have the property of 
crossing $w=-1$ on its own; the interaction does not 
solve this requirement. 

But perhaps we could reduce the amount of the shifts 
required on either end, or at least their impact, by 
{\it tilting\/} the dark energy. That is, let the 
interaction $Q$ change sign, depressing dark energy 
at high redshift and uplifting it at low redshift, or 
vice versa. 

Within a particle physics origin this seems somewhat 
unnatural. Interactions such as decays are proportional 
to $\rde$ or $\rho_m$, scattering perhaps as $\rde\rho_m$, 
but nothing expected to change sign. However, gravity 
offers a possible alternative, with for example a running 
Planck mass increasing and decreasing over the redshift 
ranges of interest (potentially adding an additional 
``why now'' issue). 

Within Horndeski gravity, the most general scalar-tensor 
theory preserving second order field equations, we can 
absorb all deviations from the Friedmann equations for 
cosmic expansion into some $\rdef$ and $\rcf$. (This of 
course does not account for the changes to the perturbative 
quantities and hence growth of structure and gravitational 
lensing. One could also absorb at the background level 
the changes to $\rho_m$ into $\rdef$, i.e.\ 
add a term $\rcf-\rho_m$ to $\rdef$ so that 
$\rho_m$ looks unchanged in the Friedmann equation, but 
this is simply sweeping the real, physical changes to 
$\rho_m$ under the rug. See \cite{bellsaw} for further 
discussion of this point.) 

For ``classic'' scalar-tensor theories such as $f(R)$ and 
Jordan-Brans-Dicke, where the only beyond general relativity Horndeski term is the 
altered coupling to the Ricci scalar, $G_4(\phi)R$, 
\bea 
\frac{2G_4}{\mpl}\,\rdef&=&\rde-6H\dot\phi\,G_{4\phi}\\ 
\frac{2G_4}{\mpl}\,\rcf&=&\rho_m\\ 
\frac{2G_4}{\mpl}\,P_{\rm de}^{\rm eff}&=&P_{\rm de}+4XG_{4\phi\phi}+2G_{4\phi}\left(\ddot\phi+2H\dot\phi\right),  
\eea 
where $\phi$ is the scalar field, $X=\dot\phi^2/2$, $H$ 
is the Hubble parameter, $M_{\rm Pl}$ the constant Planck 
mass appearing in general relativity, and subscripts $\phi$ 
denote derivatives with respect to $\phi$. Note that since 
$G_4$ only depends on $\phi$ (its  dependence on $X$ is 
disallowed by requiring gravitational wave luminality), then 
$\dot G_4=\dot\phi G_{4\phi}$. One often writes 
a time dependent ``running Planck mass'' $\mss(z)=2G_4$, and 
the property function $\alpha_M\equiv d\ln\mss/d\ln a$. 
When only $G_4$ 
enters, the braiding parameter $\alpha_B=-\alm$. 
(For completeness, the kineticity $\alpha_K=2X/(H^2\mss)$. 
See \cite{bellsaw} for expressions when other Horndeski 
terms beyond $G_4$ are present.) 

Taking time derivatives gives the continuity equations 
\bea 
\dot\rcf+3H\rcf&=&-\alm H\rcf\\ 
\dot\rdef+3H\left(\rdef+P_{de}^{\rm eff}\right)&=&+\alm H\rcf\ . 
\eea 
Therefore the results for interactions discussed in the previous sections 
apply, with here $Q=-\alm H\rcf$. 
Explicitly, 
\bea 
\wdef&=&\wde-\frac{\alm}{3}\,\frac{\rcf}{\rdef}\\ 
\wcf&=&\frac{\alm}{3}\ . 
\eea 

Thus, such modified 
gravity with $\alm>0$ (depression) or $\alm<0$ (uplift) 
will encounter the same issues already discussed. 
However, since $\alm$ is a function of time, we could 
consider the case where it passes through zero during the 
time of interest, so that a depression at one epoch turns 
into an uplift at another. 

A sign change in $\alm$, and tilting dark energy due 
to increasing and decreasing the Planck mass running, is 
not such a strange idea. Many modified gravity models 
naturally have this, and at about the right redshift 
(given that they are otherwise viable). This is seen 
for example in Fig.~3 of \cite{1607.03113}, albeit for 
a model with other Horndeski terms than just $G_4$. 

Just as before, Eq.~(\ref{eq:q3}) holds and the interaction, 
here the modified gravitational coupling, does not 
explicitly appear in $q(z)$. An advantage of tilting is 
that the dark matter equation of state will cross zero 
as $\alm$ does, presumably ameliorating issues with 
growth of structure and the ISW effect. We do note that 
the dark energy equation of state has a tilt proportional 
to $\rcf/\rdef$; this lessens the shift at low redshift 
where this ratio is less than one, but amplifies it at 
high redshift. 

Indeed one might worry that at very high redshift the 
effective dark energy equation of state $\wdef$ gets 
very large, either positive or negative. In fact this is 
not an issue since $\alm$ is time dependent and should 
go to zero as $z\gg1$ to preserve general relativity in 
the early, high density universe. Indeed, under certain 
conditions described in \cite{1512.06180,1607.03113} 
(basically the domination by a single Horndeski term 
in both the energy density and the property functions) 
-- and only those, and becoming invalid for $z\lesssim10$, 
see Fig.~3 of \cite{1607.03113} -- 
one has $\alm\propto\rdef/\rcf$ so the shift goes 
to a constant. If this did not hold (as for example at $z\lesssim10$) 
but $\alm>0$ then 
the dark energy is depressed deep into the phantom 
regime, i.e.\ the dark energy density vanishes rapidly 
into the past, similar to a phase transition case. 

At very high redshift we find that the sound speed squared 
of scalar field perturbations, which must be positive for 
stability, tends to have the form $c_s^2\propto\alm$ 
(see e.g.\ \cite{1512.06180,2003.10453}; this 
proportionality gets broken for $z\lesssim10$). 
The coefficient tends to be positive if at early times 
$\alm\sim a^s$ with $s<3.4$ and negative if $s>3.4$. 
(One can work around this depending on what bare potential 
$V(\phi)$ one takes for the dark energy.) For thawing 
fields, as we will take next, then $s=3$ at 
early times and so we need $\alm>0$. This will indeed 
depress thawing 
dark energy into phantom behavior. When $\alm>0$ 
then $\mss>\mpl$ and gravity is a bit weaker than in 
general relativity. However later when $\alm$ crosses 
zero and becomes negative (the tilt goes from depressing 
at high redshift to uplifting at low redshift), then 
gravity will strengthen. 

This may be a very attractive picture. We could take a 
standard intrinsic thawing dark energy, tilt it so that 
it is phantom at high redshift, thus causing greater 
acceleration, then the effective dark energy equation of 
state crosses $w=-1$ and $\alm$ crosses zero, and the 
intrinsic thawing equation of state is uplifted at low 
redshift, helping to avoid fine tuning issues. 
Furthermore these crossings will tend to ameliorate 
issues with the dark matter equation of state and growth 
of structure. 

The modified gravity theory has the one free function 
$G_4(\phi)$, and we want it to be nonmonotonic so that 
$2G_4/\mpl\equiv \mss/\mpl$, which is unity at very high 
redshift, then dips a little below one and later rises 
a little above one. We will not go into specific models, 
other than to note that 
$G_4(\phi)=(\mpl/2)[1+c_1\phi-c_2\phi^2]$ 
gives $\alm=(\dot\phi/H)(c_1-2c_2\phi)$, which has 
exactly the desired behavior of $\alm>0$ at high 
redshift when $\phi<c_1/(2c_2)$, and $\alm<0$ at later 
times when the field has rolled to $\phi>c_1/(2c_2)$. 
The field starts very close to $\phi_i\approx0$ to 
give general relativity in the early universe. 
Furthermore we can take the intrinsic dark energy potential 
to be of the technically natural PNGB form \cite{pngb}, 
$V(\phi)=V_\star[1+\cos(\phi/f)]$, in the thawing class. 

Given an intrinsic thawing dark energy, the stronger 
acceleration at $z\approx1$ is readily accommodated. 
Recall we want 
$q(z=1)=[1+3\wde(z=1)\Omega_{\rm de}^{\rm eff}(z=1)]/2\approx0.1\pm0.03$. 
For thawers, $\wde(z=1)$ is close to $-1$, so we need 
\be 
\Omega_{\rm de}^{\rm eff}(z=1)\approx 0.27\pm0.02\ . 
\ee 
This would depress the dark energy equation of state by 
$\approx0.9\alm$, but only uplift the dark matter equation 
of state by $\alm/3$. This may ease the constraints from 
growth of structure (especially as $\alm$ presumably goes 
through zero near this epoch). At $z=0$, to satisfy 
$q_0\approx-0.3\pm0.1$ we need $\odet^{\rm eff}\approx0.59\pm0.07$, 
assuming the bare thawer has $w_0\approx-0.9$ and is uplifted 
by $\approx0.1$ to the Gaussian process value of $\wdef(z=0)\approx-0.8$. This would give $\alpha_{M,0}\approx0.43$, 
and hence depress the dark matter equation of state by 
$\approx0.14$. 

To assess viability one would need to consider a specific 
form, and parameters, for $G_4(\phi)$, and $V(\phi)$. 
Here we simply note that the $f(R)$ subcase of having the only 
Horndeski term be $G_4$ is not suitable. 
There $\mss/\mpl=1+f_R$. We 
expect $|f_{R0}|$ to be quite near zero, perhaps 
$\lesssim10^{-6}$, but the key quantity impacting large 
scale structure observations is $f_{RR}$ or the Compton 
wavelength (squared, with respect to the Hubble scale) 
$B$ \cite{0610532}. This is closely related to $\alm$, 
\be 
\alm=B\frac{H'}{H}=-B(1+q)\ . 
\ee 
The matter power spectrum is increasingly affected 
at $k\gtrsim0.1\,h/$Mpc for $B\gtrsim10^{-5}$, hence 
$|\alm|\ll1$. 
And of course $f(R)$ gravity has solely increased 
gravitational coupling, hence $\alm<0$, and acts only as 
uplifting, without tilt.

\section{Conclusions} \label{sec:concl} 

The four conditions mentioned in Section~\ref{sec:props} 
required by taking the data at face value are extremely 
difficult to realize within the sort of dark energy 
behaviors we generally consider. Investigating  
interaction between dark energy and another component 
as a mechanism for viably crossing $w=-1$ therefore 
seems motivated. 

However, we have seen that this hope generally fails 
-- a conclusion that seems to hold fairly robustly, 
without assuming any specific form of interaction. 
The problem for both uplifting dark energy and 
depressing dark energy is that in order to fit both 
stronger acceleration in the past and weaker acceleration 
near the present, with amplitudes indicated by the data, 
one not only induces a strong shift to the dark energy density 
but also to the coupled sector, e.g.\ dark matter density 
and equation of state. These seem likely to affect 
significantly growth of structure and prove unviable. 

A potential solution is tilting dark energy, where the 
sign of the interaction changes with time. We present a 
reasonably straightforward and minimal model within 
modified gravity (using just a $G_4(\phi)$ Horndeski 
term and a standard thawing dark energy). This ameliorates 
the requirements in certain aspects but involves several 
extra parameters even for the simplest functions, and 
even then the agreement with growth of structure 
measurements is not clear. One could possibly increase 
consistency with data by including further Horndeski 
terms, but in the absence of a compelling theory the 
number of additional parameters required seems 
demotivating. 

Clear guidance from growth of structure and CMB ISW 
measurements is one path forward. Another would be the 
development of a well motivated and compactly 
predictive gravity theory. A third is the search for 
cogent reasons not to take the cosmic probe data at 
face value despite the apparent consistency of any two 
of the BAO, CMB, and supernova distance data sets.

\acknowledgments 

I thank Kushal Lodha for discussions, and Pedro 
Ferreira and William Wolf for discussions and 
hospitality at Oxford University.

\end{document}